# Performance Modeling of Distributed Deep Neural Networks

Sayed Hadi Hashemi, Shadi A. Noghabi, William Gropp, Roy H Campbell
University of Illinois at Urbana-Champaign

## ABSTRACT

During the past decade, machine learning has become extremely popular and can be found in many aspects of our every day life. Nowayadays with explosion of data while rapid growth of computation capacity, Distributed Deep Neural Networks (DDNNs) which can improve their performance linearly with more computation resources, have become hot and trending. However, there has not been an in depth study of the performance of these systems, and how well they scale.

In this paper we analyze CNTK, one of the most commonly used DDNNs, by first building a performance model and then evaluating the system two settings: a small cluster with all nodes in a single rack connected to a top of rack switch, and in large scale using Blue Waters with arbitary placement of nodes. Our main focus was the scalability of the system with respect to adding more nodes. Based on our results, this system has an excessive initialization overhead because of poor I/O utilization which dominates the whole execution time. Because of this, the system does not scale beyond a few nodes (4 in Blue Waters). Additionally, due to a single server-multiple worker design the server becomes a bottleneck after 16 nodes limiting the scalability of the CNTK.

## 1. INTRODUCTION

Intelligent systems, empowered by machine learning algorithms have dominated many aspect of our everyday life, from personal assistants on smart phones (e.g., Siri) and complex recommendation system (e.g. Netflix and Facebook), to robots and self driving cars. Nowadays, these system are expected to handle massive amount of continuously collected (TBs to PBs) [8] to train complex machine learning models with over $10^{12}$ parameters [3].

There is a large variety of machine learning algorithms, but Deep Neural Networks (DNNs) are unique among the others in performance aspects. This class of algorithms, which consists of well known simple building blocks, can achieve higher accuracy, solely by getting more computational resources. Thus, their accuracy scales linearly with the computational resources and performance of the underlying system.

DNNs have been around for decades. However, because of their excessive computational requirements wich exceeded the capabilities at that time, they did not gain much popularity. With the exponential growth in processing power and the invention of multi-core chips (both CPU or GPU), DNNs have became popular again. Currently, the rapid increase in data and problem sizes, and the need for lower latency results, has passed the limits of a single machine. Therefore, many Distributed DNN systems (DDNN) have emerged. These systems are based on Stochastic Gradient Descends (SGD), and use SGD to scale the work on multiple processors or nodes. Microsoft CNTK [14], Google DistBelief [5], and Parameter Server [8] are some the most common systems in this area.

Although DDNNs have become popular, especially in the past few years with a huge success in machine vision [7] and other areas, there has not been an extensive study on the performance benefits of these systems, their scalability, and when it is beneficial to move from a single node to a distributed environment.

In this paper, we chose CNTK, one of the most popular frameworks and evaluated the performance of the system under multiple settings. The open source CNTK code base of nearly 200K lines is actively under development [1]. This system has been used in several researches and has won some prestigious competitions. The main goal we had in evaluating CNTK was:

- Defining a performance model for "Asynchronous Stochastic Gradient Descent" (ASGD) as a baseline of the system, to understand the expected performance of the system

- Evaluating an actual implementation of ASGD and identifying how close the results are to the performance model.

- Identifying the main bottleneck of the system by further investigating cases that diverge too much from the performance model, and proposing mechanism to address these bottlenecks.

Toward this goals we ran the CNTK framework on both a 8 node cluster of beefy machines and large scale test in Blue Waters. We conducted a scalability study with varying number of nodes. Based on our results the system has very poor scalability, even though it is designed to work as a distributed system. In both test beds the system does not scale beyond 4 nodes, and even gets worse beyond that.

---
[1]http://cntk.ai/

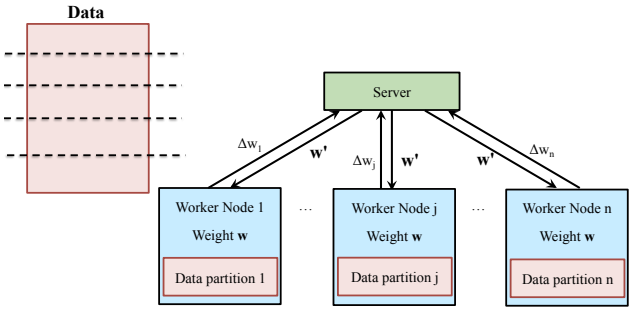

Figure 1: Architecture of Distributed Deep Neural Networks (DDNNs)

We further investigated the reason for poor scalability, by profiling and tracing the system using TAU [9]. Using this tool we realized that I/O plays a dominant role in the execution time. Since data needs to be shuffled for better accuracy and convergence, it causes sequential access to the data file, limiting the scalability of the system. Additionally, beyond a point, the single server in the system became a bottleneck, increased the wait time on the worker nodes, and in turn limited the scalability of CNTK.

We also developed a performance model for the system considering the CPU, data movement (both from memory and disk) and the network. The sysem does not perform in-line with our our initial model (Section 3) since the model assumes perfect scaling. We modified the model to take into account the sequential I/O access (Section 5.3), fitting more closer to the system.

The rest of the paper is organized as following. Section 2 gives a short background about SGD, DDNN and some core machine learning concepts. In Section 3, we develop a performance model of the system. Section 4 and 5 discuses the experimental setup and results, and also the tracing results. We conclude the paper in Section 6.

## 2. BACKGROUND

Gradient Descent, and the online version, Stochastic Gradient Decent (SGD) are one of the main building blocks for machine learning. In these optimization algorithms, each data point is represented as a vector $\vec{d}$ of features. The goal is to optimize a object function (in DNN, a neural network) by finding optimal parameters (in Neural Network weights) $\vec{w}$ of size $n$ based on a set of data points and their expected values.

In SGD, each data point is evaluated in the model (with current parameter $\vec{w}$), to produce a result. The result is compared to the expected result which then updates the parameters accordingly ($\vec{w'}$). The difference in the parameters, $\triangle w = \vec{w'} - \vec{w}$, is used to minimize the error of the model through a process called "back-propagation". This process is repeated several times for each data points until a certain level of optimization (in DNN accuracy) is reached.

In the original SGD model, the $\triangle w$ is applied to $\vec{w}$ per each data point. However DNNs use a slight variant in which a batch of datapoints are processed before the $\triangle w$ is applied. This improves both performance and accuracy of the neural network. The batch size is configurable, and as shown in Figure 4 larger batch sizes can improve performance up to a point.

Distributed Deep Neural Networks (DDNNs) work based on Stochastic Gradient Descent (SGD). In majority of these system the size of the input data is massive, and beyond the capability of a single machine (in a timely manner). Thus, these system chunk the input data into multiple partitions spread across many machines, so called workers. Each machine works on one partition of the data in parallel with other workers, as shown in Figure 1.

By partitioning the data each worker can reach a partial suboptimal result, since it does not have a complete view of the data. Thus, all these workers send an updated delta of their weight ($\triangle w$) to a central server. The server aggregates these updated weights into a new weight (w'), and propagates the new weight back to the workers.

For further performance improvement, the DDNN systems are designed in a asynchronous manner, with almost no barrier between workers. Each worker independently sends the weight updates ($\triangle w$) to the server, and does not wait for others. Also, the server asynchronously aggregates the weight updates. The only main barrier is when one worker has gone through all its data, which it waits for others to reach the same point. Note that in SGD the whole data is iterated over multiple times, typically until a max has reached or the algorithm has converged. However, since the data is large this barrier should not be reached frequently.

Note that these are how the systems are designed, but as we will see not all these designs perform as well as expected in reality.

## 3. PERFORMANCE MODEL

As the first step we generated a performance model for the system, so that we can have a basic understanding of the expected behavior of the system. Since the system is very complex with tens of thousands of lines of code, it was not feasible to study the system in a instruction level. Thus, we built our performance model in higher level by treating the system as a black box with a specific set of properties. We do not include all the details of the implementation, and the main considerations we have are designs discussed in Section 2.

We define the parameters of the performance model as shown in Table 1. We gather these parameters by running performance test such as Stream (for memory bandwidth) or iperf (e.g., disk bandwidth) [12] , by using specifications of the system (e.g. disk bandwidth), by setting them in the experiments (e.g., $epoch, d, w$), or by running the test on a single node and extracting the values, e.g. $c_p, c_b$.

We model the system from the point of view of each **worker**. Also, since the data is large we do not consider the cache in our modeling, and assume data always read from memory.

**Computation:**
With these assumptions and parameters, on each iteration (also call epoch), each worker processes $\frac{d}{w}$ data points. The processing of $n$ data samples once would be

$$T_{processing}(n) = c_p \times n \qquad (1)$$

Note that $c_p$ is not a constant in general since it depends on several parameters such as $n_f$ (number of features), and the neural network structure. However, we run all our experiments with a fixed $n_f$ and network to keep the model simple. This allows us to assume for simplicity that $c_p$ is constant throughout all test.

| Parameter | Definition |
|---|---|
| $w$ | number of workers in the system |
| $b$ | batch size |
| $n_f$ | number of features in the datapoint vectors, i.e. $|\vec{d}|$. Each datapoint vector of size $n_f$ of double entries. |
| $n_w$ | number of dimensions in the weight vectors, i.e. $|\vec{w}|$. Weight is a vector of size $n_w$ of double entries. |
| $d$ | number of data points in the input data |
| $epoch$ | number of iterations happened until convergence |
| $c_p$ | time spent for processing one datapoint. Since in all test the data had the same size of $n_f$ with the same cardinality, we assume all processing each datapoint takes a constant time $c_p$ |
| $c_b$ | time spent for batching data and sending it to the server. In this process a $\triangle w$ is sent to the server, and this overhead is almost constant regardless of the batch size. |
| $c_u$ | time spent on server to aggregate $\triangle w$ received from workers. |
| $r_d, w_d$ | time spent to read/write one byte from/to the disk. We assume $r_d = w_d$ |
| $r_m, w_m$ | time spent to read/write one byte from/to the mem. We assume $r_m = w_m$ |
| $r_{net}$ | time needed to transfer one byte over the network, i.e, 1 over the bandwidth of the network |

Table 1: Parameters used in performance model

In addition to the processing, each worker has to prepare and send an update $\triangle w$ for each batch of data. Thus, the total update overhead would be

$$T_{update}(n) = c_b \times \frac{n}{b} \quad (2)$$

Overall total computation time for an iteration would be

$$T_{computation}(n) = T_{processing}(n) + T_{update}(n) \quad (3)$$

$$T_{computation}(n) = c_p \times n \times (1 + \frac{1}{b}) \quad (4)$$

As shown in this formula, we expect smaller batch sizes to increase the overhead. Also, we expect the computation to scale linearly with adding more workers from the computation perspective since it reduces the number of samples each worker has to compute. Thus, by doubling the workers we expect the execution time to be cut in half.

**Data:**
Each worker processes $\frac{d}{w}$ of the data points. This data has to be loaded from disk into memory at least (and commonly) once. Data points are read from memory in batch and being computed once in each iteration, and never accessed until the next. Assuming data points and weights are vectors of *doubles*, since $n$ data points are loaded once from disk, we would have:

$$T_{disk}(n) = r_d \times 8 \times n_f \times n \quad (5)$$

In the computation phase, each batch of data points is read from the memory. In addition, weight vector needs to be read and written during an iteration. We assume the weight vector can be kept in the cache along with the batch during an iteration. Thus for the data movement we have:

$$T_{memory}(n) = r_m \times 8 \times n \times (n_f) + (r_m + w_m)\frac{8 \times n \times n_w}{b}$$
$$= 8 \times n \times (r_m \times (n_f) + (r_m + w_m)\frac{\times n_w}{b}) \quad (6)$$

However, if the weight vector does not fit in the cache (as we will see later in Section 4.4) the data movement time increases rapidly.

**Communication:**
Many messages are sent around in this framework. All the communication is done using MPI Isend and Irecv. Thus, the computation should not block on the communication. For each batch size, each worker sends the $\triangle w$ vector to the server. We use the simpler version of $s + rn$ model by ignoring $s$ here. Thus we have:

$$T_{send}(n) = r_{net} \times 8 \times n_w \times \frac{n}{b} \quad (7)$$

The server spends $c_u$ to aggregate the updates and broadcast the result back to workers. Each worker also receives a new weight $w'$ for each batch of data. Given that the size of $w'$ is the same as $\triangle w$, and they are sent per each data batch, the total time would be:

$$T_{network}(n) = 2 \times T_{send} + w \times c_u \times \frac{n}{b}$$
$$= \frac{n}{b} \times (16 \times r_{net} \times n_w + w \times c_u) \quad (8)$$

**Overall:**

The network is done in parallel with the computation. But, the data transfer and computation are not pipelined by the developer in the system.

Therefore the overall time to process $d$ data points with $w$ workers is:

$$T_{total}(d, w) = T_{disk}(\frac{d}{w}) + epoch \times$$
$$max(T_{network}(\frac{d}{w}), T_{computation}(\frac{d}{w}) + T_{memory}(\frac{d}{w}))$$
$$= \frac{1}{w} \times (T_{disk}(d) + epoch \times$$
$$max(T_{network}(d), T_{computation}(d) + T_{memory}(d)) \quad (9)$$

## 4. EXPERIMENTAL RESULTS

We conducted two different sets of experiments on two different testbeds. First, we micro benchmarked the system on a small cluster (Mustang) in a blackbox fashion. Our goal was gasp a base understanding of the performance of the system, and find the limitations. Next, we extends the experiment by conducting a more detailed benchmark and tracing of the system on Blue Waters at larger scale (Section 5).

## 4.1 Goals

The main goal of DDNS is to partition the data and computation on multiple machines, and get better performance by doing so, or tackle large problems that could not be solved on a single machine. In either case, the dominant factor is how well the system scales by adding more resources. Thus, the focus of our experiments is to study the scalability of the system. The main goals we had by conducting these experiments were:

1. Studying the scalability of the systesm to see if it scales linear, sub-linear or super-linear.

2. Identifying where the system cannot scale beyond.

3. Discovering the reason and main bottleneck for not limited scalability in the system.

4. Trying to improve the scalability by resolving the bottleneck.

In addition to the scalability study, we perform studies to see the effect of tunable parameters in the system. In detail, we focus on how the batch size effects performance and scalability of the system.

## 4.2 Experimental Setup

### 4.2.1 Test Bed

We conducted our experiments on two testbeds, discussed below.

**Mustang:**
Mustang is a distributed 7 node cluster of beefy machines. All nodes are placed in a single rack with a Top of Rack switch connecting them together. We chose this cluster since we knew exactly how nodes are connected, and with a single rack the effect of the location of nodes was minimal. The most important information about this testbed is shown in Table 2.

**Bluewaters:**
Blue Waters is a HPC super computer with more than 700K nodes. Using Blue Waters we were able to test the system in larger scale, and with InfiniBand network. However, we did not have high control on where nodes were placed and how congested the links were (since many other applications are running on the system). The main information about this testbed is shown in Table 3.

### 4.2.2 Workload

**Network:**
We modified an existing example in CNTK (Examples/Image/MNIST/Config/01_OneHidden.ndl). This network has three layers:

1. An input layer with dimensions equal to number of input features (in our data $d_i = 2$)

Table 2: Mustang cluster information

| Nodes | 7 |
|---|---|
| CPU | 2 × Intel(R) Xeon(R) CPU E5-2660 |
| Clock | 1.20 - 3.00 GHz |
| Cores | 2 × 8 physical, 2 × 16 logical |
| L1i | 32 KB |
| L1d | 32 KB |
| L2 | 256 KB |
| L3 | 20480KB |
| Memory | 128GB |
| HD | 256GB SSD |
| Network | 10 Gbps Ethernet |
| Compiler | GCC 4.9.3 |
| MPI | OpenMPI 1.10.2 |
| BLAS | OpenBLAS 1.2.20110419-10 |

Table 3: Bluewater cluster information

| Nodes | 64 |
|---|---|
| Cores | 16 × 64 |
| Clock | 2.3 GHz |
| Cores | 2 × 8 physical, 2 × 16 logical |
| L1i | 64 KB |
| L1d | 16 KB |
| L2 | 2048 KB |
| L3 | 6144 KB |
| Memory | 512GB |
| Network | InfiniBand |
| Compiler | GCC 5.3.0 |
| MPI | Cray MPICH 7.3.0 |
| BLAS | Intel MKL 15.0.3.187 |

2. A hidden layer with two different settings for dimension ($d_h$): 10K and 100K

3. A label layer with dimension equal to different possible label values (in our example $d_o = 2$)

Therefore, we will have:

$$\begin{aligned} n_w &= d_i * d_h + d_h * d_o \\ n_w^{10K} &= 40K \\ n_w^{100K} &= 400K \end{aligned} \quad (10)$$

in both setting, weights can fit in the cache. More details of the workload is discussed below.

**Data:**
We generate a synthetic random dataset for experiments. The dataset consists of 16384 data points, each with two features of type double and an integer label. The data is store as a tab separated values text file.

**Config:**
We slightly modified an included example in CNTK (Examples/Other/Simple2d/Config/Multigpu.cntk). In this configuration, data is read by UCIFastReader module. This module reads data as tab-separated values files sequentially, then may randomize the records.

### 4.2.3 Measurements

In all experiments, the main focus was execution time of the workload. We started our measurement using the *perf* command. We also measure the usage of various resources (CPU, GPU, network, disk, I/O, etc.) both on the workers and the server. For this purpose we use a combination of multiple tools including:

1. **dstat**: dstat [6] is a tool to designed to continuously monitor many resources including CPU, memory, disk and network, while imposing little to no overhead. In our experiments we gathered metrics every second and computed the max/average over the time.

2. **NVIDIA System Management Interface**: dstat does not provide any information regarding GPUs and how well they are performing. Thus, we used nvidia-smi [1], a tool provided by NVIDIA which provides continuous detailed stats.

3. **iperf**: iperf is a tool to actively measure maximum achievable bandwidth of the network. We used this tool to gather information about the network utilization.

Additionally, when we found that the system does not scale beyond a certain point, we used some tracing and profiling tools to further investigate the cause. After examining a few different approaches and tools, we used the PMPI interface of CNTK to generate profiling and tracing reports. More details in §5.

## 4.3 Scalability Study

We measured the total execution time while adding more nodes to the cluster for various batch sizes, shown in Figure 2. As seen, the system scales well from one node to two nodes. However, after that not only the execution time does not decrease proportionally, it even increases in many cases. In some cases, e.g. batch size 1000 and 5, it does not even benefit from going from one node to two nodes.

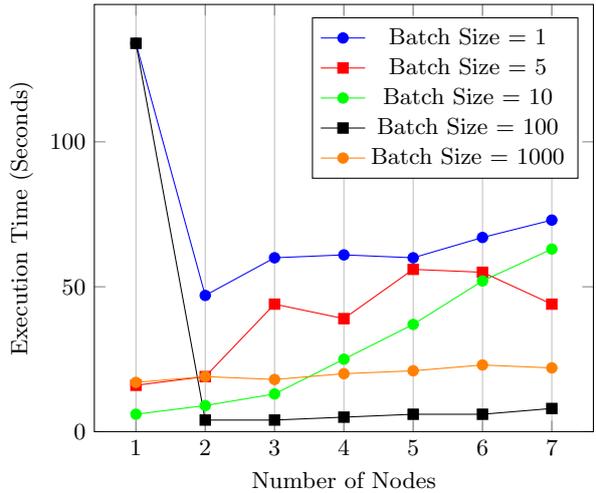

Figure 2: Scalability study with various batch sizes on Mustang test bed. These experiments use $\#Neural = 2 + 10K + 2$

We repeated the test on Bluewaters with same number of neurons and 10 times more. As seen in Figure 3, the same behavior is observed. The system scales up to a few nodes, then it starts flattening and even getting worse. Also, for $> 16$ nodes, the difference between computing a small (blue circles) and large (red squares) neural network is not significant, even though the large network is almost 10 times larger. This indicates that the majority of the execution time is spent on extra overheads, rather than computation on the model.

Additionally, we measured the resource utilization (CPU, disk, network) on the workers, and found that none of them are close to saturation. The utilization was $< 30\%$ in all test for all resources. This indicates that the poor scalability is either because the server gets overwhelmed with requests, or there are some barriers and synchronization diminishing the benefit of more parallelism. We have further analyzed the reason for this poor scalability by tracing and profiling the system in §5

Although we expect better performance by increasing the batch size (based on out performance model in Section 3), our results do not always follow this pattern. This is because of ignoring the effect of cache in our model, discussed in more detail Section 4.4

## 4.4 Effect of Batch Size

We ran the workload on the Mustang cluster with various batch sizes and nodes, shown in Figure 4. As seen, increasing the batch size decreases the execution time up to a point, and then jumps up and starts decreasing again. This pattern does not follow our performance model designed in Section 3. The main reason is that the model does not include the effect of cache. For small batch sizes ($\leq 64$), all the datapoint and the weights fit into cache. The size of the datapoint and weight is:

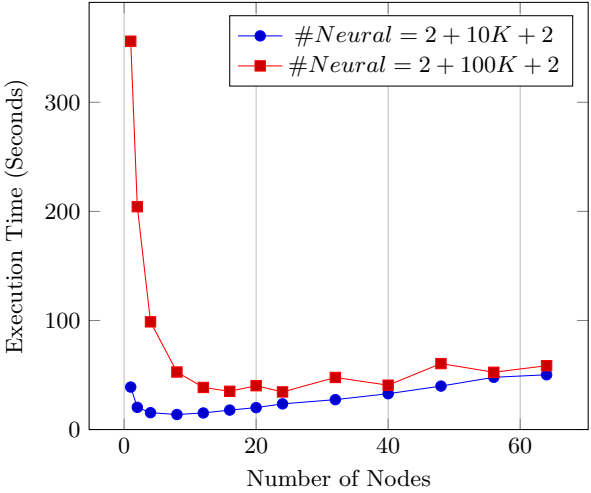

Figure 3: Scalability study with various hidden network dimensions on Bluewaters test bed. These experiments use $batchsize = 256\ samples$

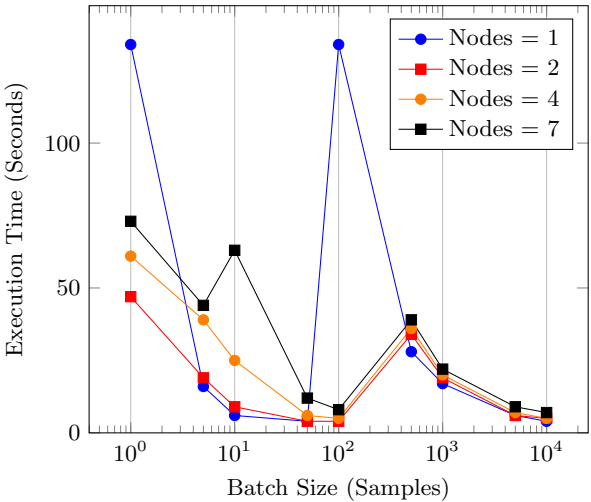

Figure 4: Effect of batch size on execution time. Results were gathered from Mustang with varying number nodes and 24B datapoints

$$\begin{aligned}datapoints =&(features+labels) \times size(double) \times batchSize\\ =&(2+1) \times 8 \times 64 = 2.4KB\\ weight =&40K \times size(double) = 320KB\end{aligned}$$
(11)

which clearly fits into L2 and L3 cache. Thus, by loading the datapoint the weight vector does not get evicted from cache and does not need to be reloaded over and over again. However, for lager batch sizes the data no longer fits into cache and there is sudden jump of ≈100x, which is a reasonable increase from cache to memory.

The peek for one node is higher than >2 nodes, while afterward the data points fall on each other. We suspect a similar peek should happen for >2 nodes somewhere in the range of [128-256] batch size. However, we do not have enough data points in that range to capture the peek.

## 5. TRACING AND PROFILING

As shown in Section 4.3, our experiments did not scale beyond a few nodes (e.g. 2 nodes for a small neural network on the Mustang test bed). The main suspicion we had was that long barriers and waits and a single bottlenecked server were the issue. To get more detail information we investigate two different approaches: full instrumentation, and MPI PMPI interface.

In our experiments with full instrumentation, Score-P [2] , added huge overhead (> 10x slower) making tracing results unusable for performance modeling.

On the other hand, since CNTK was using MPI as its communication mean, we used MPI monitoring tools to profile and trace our experiments. These tools take advantage of the PMPI interface to monitor with minimum overhead. We tried many tracing tools, but faced a few challenges in using many of them. For instance, CNTK uses MPI3 with *multiple thread* mode, with is not fully supported by VampirTrace [10] (included in OpenMPI). Also, CNTK does not properly terminate the MPI (using either *MPI_Finalize*, or *MPI_Abort*) in all the cases. This misbehavior interrupts the functionality of MPI Parallel Environment [4] and eztrace [13].

The only reasonably working tool we found were *Integrated Performance Monitoring* [11] (IPM) which is a very light weight profiling tool, and *Tuning and Analysis Utilities* (TAU) which supports MPI event-based sampling tool.

Additionally, since we wanted to analyze the system in larger scales (beyond 8 nodes) we ran the profiling and tracing experiments using Blue Waters. The result from the Mustang cluster was the same, and we have omitted them due to lack of space.

### 5.1 Profiling

We used IPM profiling tool to further analyze scalability of CNTK. This tool measures the ratio of computation time, barrier time (i.e., time spent in MPI_BARRIER), and wait time (time spent in MPI_WAIT) to the total execution time. We measured this ratio for each node in a 64 node test running on Blue waters.

As shown in Figure 5, on average > 50% of the time is spent in MPI barrier among nodes with a very high standard deviation. Additionally, the barrier time is the most in node 0, linearly increasing to node 63. This indicates that there is part of the execution done sequentially, since the barrier time is increasing linearly. In detail, node 0 computes and waits on the barrier, then node 1 computed and joins waiting on the barrier, and this pattern continues until the last node.

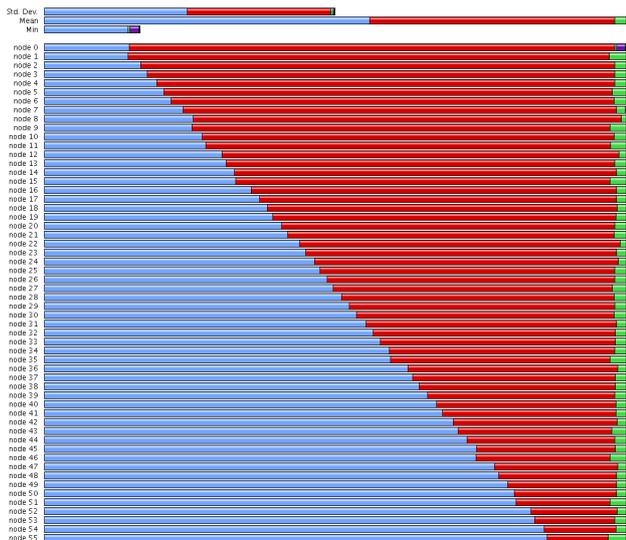

Figure 5: Ratio of computation (blue bar), barrier time (red bar), wait time (green bar) for each node on 64 node test on Blue Waters. These results are gathered by IPM profiling tool.

Table 4: Tracing key codes and descriptions.

| Key | Description |
|---|---|
| ■ | .TAU application |
| ■ | FLUSH |
| ■ | MPI_Allreduce() |
| ■ | MPI_Barrier() |
| ■ | MPI_Comm_rank() |
| ■ | MPI_Comm_size() |
| ■ | MPI_Init_thread() |
| ■ | MPI_Irecv() |
| ■ | MPI_Isend() |
| ■ | MPI_Wait() |
| ■ | MPI_Waitall() |
| ■ | MPI_Waitany() |
| ▽ | Preview_Event |
| ▽ | Message size for all-reduce |
| ▽ | Message size received from all nodes |
| ▽ | Message size received in wait |
| ▽ | Message size received in wait : .TAU application => MPI_Wait() |
| ▽ | Message size received in wait : .TAU application => MPI_Waitany() |
| ▽ | Message size sent to all nodes |

## 5.2 Tracing

We used TAU to gather detailed traces of the system. We ran the test on Blue Waters and Mustang. Since we went to larger scales in Blue Waters and Mustang results were similar, we have only presented results from Blue Waters. Table 4 describes the legends for all results gathered from TAU.

### 5.2.1 Total Execution Trace

Figure 6 shows the detail trace of the whole execution for 2 and 64 nodes. The blue arrow shows the initialization time, i.e., the time spent to read the chunk of data the node is in charge of, setting up the environment, and starting processing on the first batch of data. The read arrow shows the computation which includes processing a batch of data, sending an update $\triangle w$ to the server, receiving the new weight $w'$ and a small barrier at the end of one iteration over the whole dataset.

As shown the total execution time increases $> 3x$ from 16 second in 2 nodes to 50s in 64 nodes. Also, the more nodes we have in the system, the ratio of initialization increases from 25% (2 nodes) to $> 90$ % (64 nodes). This shows that the dominant factor in the execution time is the initialization phase and this is the main reason the system does not scale well.

The performance model (Section 3), does not model the initialization phase properly. We have updated the performance model accordingly in Section 5.3.

In addition, we have analyzed each of the initialization and computation phase in more detail in Section 5.2.2 and 5.2.3.

### 5.2.2 Initialization Phase Trace

We further analyzed just the initialization phase for both 2 and 64 nodes, as shown in Figure 7. The main cause of the long initialization phase, specially for larger scales, is a barrier on the execution (the light blue boxes). This barrier linearly decreases from the first node to the last node. This indicates sequential access for a some part of the initialization, instead of parallel computation. In detail, first node does the computation and waits for the rest. When node 1 is done, only then node 2 enters that section. We believe this sequential access is due to poor I/O performance in the initial distribution of the data (each node receiving its chunk of data).

We looked at the source code, and figured the data is not partitioned by splitting the data file into chunks (as shown in Figure 1), since this would lead to poor accuracy in DDNNs. Instead data has to be shuffled around, with each node receiving $\frac{1}{n}$ of the data. The way this is done in CNTK is that for each node, out of the whole file, randomly some lines are assigned to that worker. This causes long barriers and also random access to disk with many disk seeks.

As part of future work, we plan to further investigate this issue and work on solutions to remove the long linearly increasing barriers, parallelize the I/O, and improve the I/O performance.

### 5.2.3 Computation Phase Trace

We also, analyzed the computation phase in detail, shown in Figure 8. The white lines indicate a communication, the dark green boxes are MPI_SENDs and the orange boxes are MPI_WAITs. Note that the server is also running on the first node.

As shown, the workers asynchronously send an update to the server, roughly around the same time. When the server receives all the updates, it will compute a new weight and send it back to the workers. However, the wait time increases when going from 2 nodes to 64 nodes. This shows the server becomes overwhelmed with more workers, becoming the bottleneck of the system. As part of future work we plan to further investigate this issue to reduce the load on the server.

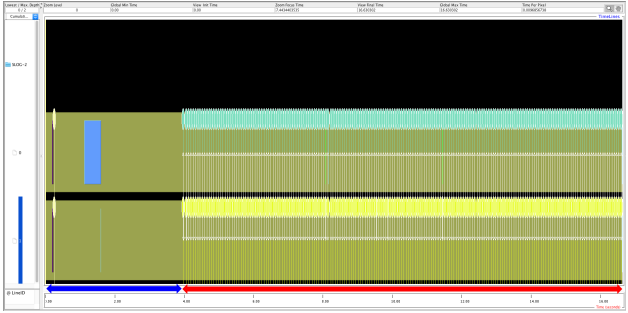
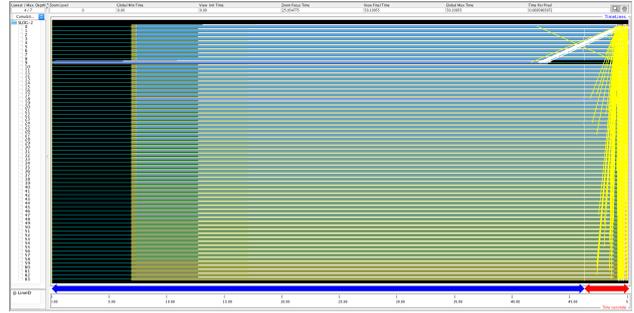

(a) Nodes = 2            (b) Nodes = 64

Figure 6: Detailed tracing of a test with 64 and 2 nodes running on Blue Waters. The trace was gathered by TAU. The blue arrow indicates the initialization and the red arrow indicates the computation time. Legends are described in Table 4.

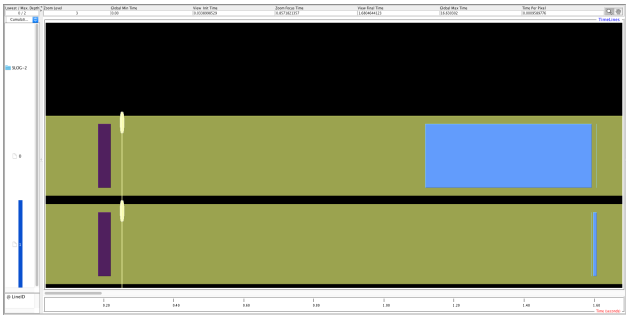
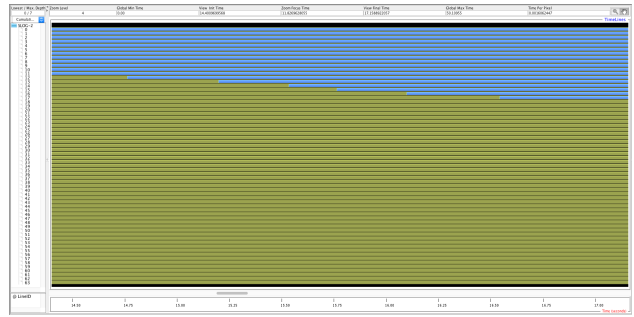

(a) Nodes = 2            (b) Nodes = 64

Figure 7: Detailed tracing of the initialization phase with 64 and 2 nodes running on Blue Waters. The trace was gathered by TAU. Legends are described in Table 4.

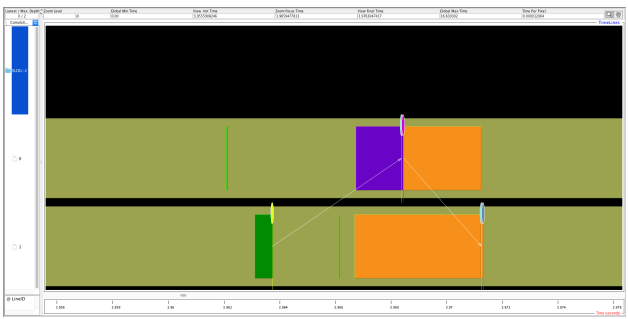
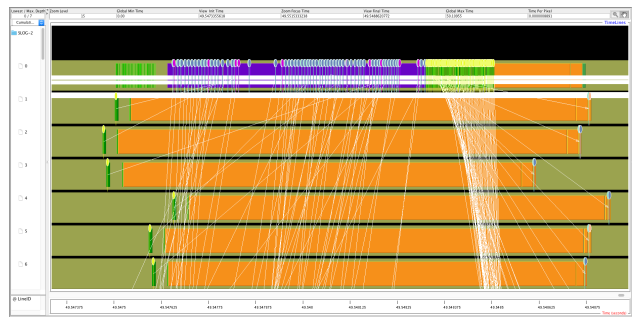

(a) Nodes = 2            (b) Nodes = 64

Figure 8: Detailed tracing of the computation phase with 64 and 2 nodes running on Blue Waters. The trace was gathered by TAU. Legends are described in Table 4.

In addition, we have analyzed the trend of both initialization and computation phase with adding more nodes in Section 5.4

## 5.3 Updated Performance Model

Based on our results, the performance model needs to be updated. In particular we update $T_{disk}$ and $T_{total}$.

**Disk:**
Unlike our first assumption that each process will read only part of the dataset out of the disk, the processes reads the whole dataset, but keeps only a fraction in the memory.

Clearly this behavior of initialization phase is not scalable. Even worse, on Bluewaters with shared I/O node between processing nodes, initialization phase takes longer with more nodes. Therefore The new model on the Bluewaters is:

$$T_{disk}^{new}(w, n) = w \times T_{disk}(n \times w) \quad (12)$$

**Total:**
The other aspect we had missed in the performance model is the minimum achievable computation phase time ($T_{computation\ phase}$). Therefore we have:

$$T_{total}^{new}(d, w) = T_{disk}^{new}(w, \frac{d}{w}) + epoch \times \\ min(max(T_{network}(d), T_{computation}(d) + T_{memory}(d)), \\ T_{computation\ phase})$$

## 5.4 Phase By Phase Scalability

As shown in Section 5.2.1, The initialization phase is the major cause of poor scalability. We studied each of the phases separately with varying number of nodes to find the scalability trend, the point where the system does not scale beyond, and the potential performance gain by improving each of the two phases. We performed the same analysis as Section 5.2.1 using TAU tracing on varying number of nodes. Figure 9 shows the result of this experiment and the expected performance based on updated model (Section 5.3).

As shown, The initialization phase (which is dominated by the I/O) increases linearly by adding more nodes. As discussed in Section 5.2.2, this is due to sequential I/O access in the initialization phase.

Additionally, the computation decreases almost linearly (i.e., by doubling resources, time drops by half) up to some point, 16 nodes in our case, and then starts increasing. The reason for the increase is that the server becomes saturated with 16 nodes and beyond 16 it negatively impacts performance. This shows that even if the initialization phase is completely parallelized, the system still won't scale beyond 16 nodes.

In total, since initialization is the dominant factor, the overall scalability of the system is fairly poor (not scaling beyond a very few nodes), with a great space for potential improvement.

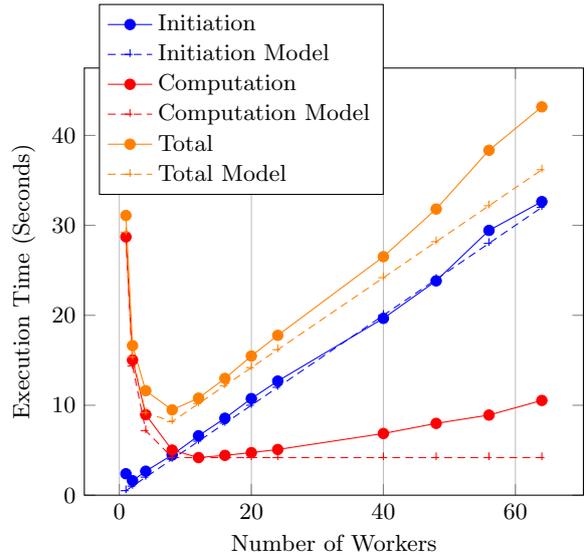

Figure 9: Scalability study and modeling of phase by phase (initialization and compuataion) and total execution with varying number of nodes. The experimental results were gathered by TAU's tracing tool on Blue Waters with $batchSize = 256$ and $\#Weights = 40K$

### 5.4.1 Perfromance Model

We also compared our updated performance model (Section 5.3) to the experimental results, as shown in Figure 9. As it can be seen our updated model is a close fit to the results gathered. With the update for the I/O access, the model is a very close match to the initialization phase. The small difference in the tail of the computation and the computation model is because the model does not incorporate the server becoming a bottleneck. The model simply assumes a limit on the server capacity, but does not consider any additional overhead for scaling beyond that point.

## 6. CONCLUSION

In this paper we studied CNTK, one of the major Distributed Deep Neural Network frameworks. Our main focus was the scalability of the system. Towards this goal, we built a performance model closely fitting the real world experiments. Additionally, using micro benchmarks, we evaluated the scalability on two different test beds. Based on our results, we found CNTK does not scale beyond a few nodes (4 nodes on Blue waters).

Using detailed profiling and tracing tools we realized the poor scalability is due to two reasons: 1) poor I/O performance which became the main overhead in the computation 2) having a single server which became overwhelmed and the bottleneck in the execution.

As part of future work we plan to investigate more on how to mitigate the poor scalability issue. First, we plan to improve the I/O performance by utilizing parallel access and reducing the disk seeks. Second, we plan to reduce the load on the single server to be able to scale to larger sizes.